\documentclass{article}

\usepackage{times}
\usepackage{enumitem}
\usepackage{graphicx} 
\usepackage{subfigure} 

\usepackage[]{natbib}

\usepackage{algorithm}
\usepackage{algorithmic}
\usepackage[utf8]{inputenc}

\usepackage{amsthm}
\usepackage{amsmath}
\usepackage{bm}
\usepackage[english]{babel}

\def\app#1#2{%
  \mathrel{%
    \setbox0=\hbox{$#1\sim$}%
    \setbox2=\hbox{%
      \rlap{\hbox{$#1\propto$}}%
      \lower1.1\ht0\box0%
    }%
    \raise0.25\ht2\box2%
  }%
}
\def\approxprop{\mathpalette\app\relax}

\usepackage{hyperref}

\theoremstyle{definition}
\newtheorem{definition}{Definition}[section]

\usepackage[accepted]{icml2017}
\usepackage{dblfloatfix}
\usepackage{graphicx}
\graphicspath{ {images/} }

\icmltitlerunning{Latent Racial Bias - Evaluating Racism in Police Stop-and-Searches}
\begin{document} 

\twocolumn[
\icmltitle{Latent Racial Bias - Evaluating Racism in Police Stop-and-Searches}




\begin{icmlauthorlist}
\icmlauthor{Akbir Khan}{cam}
\end{icmlauthorlist}

\icmlaffiliation{cam}{University of Cambridge, Cambridge, United Kingdom}

\icmlcorrespondingauthor{Akbir Khan}{ak2062@cantab.ac.uk}

\icmlkeywords{boring formatting information, machine learning, ICML}

\vskip 0.3in
]



\printAffiliationsAndNotice  

\begin{abstract} 

In this paper, we introduce the \textit{latent racial bias}, a metric and method to evaluate the racial bias within specific events. For the purpose of this paper we explore the Home Office's recent live feed and dataset of stop-and-search incidents. We explore the racial bias in the choice of targets, using a number of statistical models such as  graphical probabilistic and TrueSkill Ranking. Firstly, we propose a probabilistic graphical models for modelling racial bias within stop-and-searches and explore varying priors. Secondly using our inference methods, we produce a set of probability distributions $\beta_{k}$ for $k$ racial/ethnic groups based on said model and data. Finally, we produce a set of examples of applications of this model, predicting biases not only for stops but also in the reactive response by law officers. 

\end{abstract}

\section{Introduction}

Stop and searches have been a form of policing since the Vagrancy Act 1824, acting as a policing tool which provides broad and discretionary power to Police officers. It has come under regular legal, political and societal scrutiny due it's reliance on the discretion of police officers. However it has also been praised as it has not only combated crime but also for being preventative. Since its formal introduction as the 'sus' laws of 1824, stop and searches laws have been subject to many reforms, due to the unfair implementation within the black community (see 1981 riots in Brixton Londonm Handsworth, Toxteth Liverpool, Southall \& Moss Side Birmingham). 

Stop and search is a fundamentally subjective tool, predicated on the trust that an officer's views are wholly unbiased and prejudice. The vast majority of search power now require 'reasonable suspicion' of an offence to be present \cite{stop}. These offences are categorised into 4 groups: 
\begin{itemize}[noitemsep]
  \item illegal drugs
  \item a weapon
  \item stolen property
  \item something which could be used to commit a crime, such as a crowbar
\end{itemize}

Alternatively, during times of extraordinary occurrences, a commissioner may designate specific areas as having `reasonable suspicion', these however are infrequent. 

For this project, we aim to explore the racial\footnote{In this paper, we do not make distinction between the terms "ethnicity" and "race", which are used to refer to people who share common facial features that perceptually distinguish them from members of other ethnic groups.} bias within the discretion of police officers within the stop-and-search procedure. We produce a probabilistic graphical model to model such a procedure, which is parameterised by a learnable latent variable $\beta$, which can be interpreted as the racial bias. We explore this bias not as fixed set of values, but as a probability distribution encoding modal, mean and variance of results. Applying Bayesian inference we are able produce probability distributions for the racial bias $\beta_{k} \sim P(\beta_{k})$ for an individual of a specific race $k$. 

In this paper, we hope to demonstrate how probabilistic machine learning can be used to infer hidden variables within opaque processes, thus shedding light on them and the potential inefficiencies which may exist.

In Section 2, we include an introduction to the mathematical tools needed to model the process, such as Bayes rule, Graphical models and TrueSkill Ranking. In Section 3 we apply the aforementioned methods to produce the key model for our data. These models are then evaluated in Section 4 and  a discussion is included in Section 5.

\section{Background}

\subsection{Bayesian Inference}
A probability distribution $P(x_{1}, ... x_{n})$ can be interpreted as providing a statement of certainty of an event occurring given a number of factors $x_{i}$. It is usually difficult to describe $P(x_{1}, ... x_{n})$ analytically, however it is significantly easier to describe the relation between individual factors with others. To express this dependency we use conditional probability distributions $P(x_{1}, ..., x_{n-1}|x_{n})$, which describe the conditional information between an observation $x_n$ and the certainty of the statement. Thus we can describe conditional probability distributions between factors, can be seen as the building blocks for modelling larger more complex distributions, e.g $P(x_{1}, ... x_{n}) = P(x_{1},..., x_{n-1}|x_{n}) P(x_{n}$). 

We are particularly interested in discovering if certain variables, $y_{i}$ have any influence upon an event given we have data of the event and it's occurrences. We require a form of statistical inference which allows us ability to infer pre-conditions given the outcome of events. To this end we utilise we use Bayes' rule:
\begin{equation}
P(A|B) = \frac{P(B|A)P(A)}{P(B)},
\end{equation}
where $A$ and $B$ are events and $P(B) \neq 0$. 

Bayes rule in itself, is for an individual event. For a larger statistical model, we attempt to produce a model P, upon a set of observed points $\mathbf{X}$, parameterised by $\theta$ such that $x \sim P(x|\theta)$. There may also exist some abstraction or hyper-parameters $\alpha$ which determines characteristics of $\theta$ such that $\theta \sim P(\theta|\alpha)$; this is often referred to as the prior distribution.

To evaluate the probability of the data given our model, we calculate the likelihood as $p(\mathbf{X}|\theta)$.  We can also evaluate the  probability of the data over all possible models with this set of $\theta$ as the marginal likelihood, given by $\int_\theta p(\mathbf{X}|\theta)p(\theta|\alpha)d\theta = p(\mathbf{X}|\alpha)$. Combining these three measures: prior, likelihood, marignal likelihood with Bayes rule, we can calculate a distribution of parameters:

\begin{equation}
p(\theta|\mathbf{X},\alpha)=\frac{p(\mathbf{X}|\theta)p(\theta|\alpha)}{p(\mathbf{X}|\alpha)}
\end{equation}

This provides the unique ability of not only telling us the most likely variable $y_{i}$ is within our model, but also shows us the entire probability distribution for this (thus we also get a measure for it's variance, skewness and shape). 

\subsection{Graphical Models}
Graphical models (GMs) are an effective mathematical tool for the depiction of such relationships between probability distributions. Extensions of Bayesian networks, they depict dependency relations between key probability distributions \cite{nielsen2009bayesian,lauritzen1996graphical}. However they also encompass additional assumptions regarding the structure of these probabilities such as chain dependencies and local/global markov properties. 

Each class of GM is a particular union of graph and probability constructs and details the form of independence assumptions represented. For the purposes of our study, we choose to utilise Factor Graphs (FGs):
\theoremstyle{definition}
\begin{definition}{\textit{(Factor Graph)}}
Given a function,
\begin{equation}
f(x_1, ..., x_n)= \prod_{i} \psi_{i} (\chi_I)
\end{equation}
A FG has a node (represented by a square) for each factor $\psi_{i}$, and a variable node (represented by a circle) for each variable $x_{j}$ . For each $x_{j} \in \chi_{i}$ an undirected link is made between factor $\psi_{i}$ and variable $\chi_{j}$.
\end{definition}
FGs are well suited to inference algorithms, as they express the factorisation structure of a model, allowing us to inspect conditional distributions. Factor graphs are able to propagate information through subgraphs effectively based on the markov chain assumptions. Ultimately this lends itself for applications of Bayes rule and generating distributions for individual factors. 

FGs allow us to quickly calculate an optimised form of gibbs sampling (described below) called message passing. This is guaranteed to enable us to calculate any marginal $p(x_{i})$ using a number of summations which scales linearly with the number of variables in the tree.

\subsection{Gibbs Sampling}
The computations required within Eq 2 require evaluation of complex integrals, this is common with marginal distributions. In general cases, the tractability of such integrals is rarely guaranteed, as is the practicability of approximation by an integration by a sum of products. Thus, for a generic function $\phi(x)$ and distribution $p(x)$, we apply a monte carlo approximation:
\begin{equation}
\int \phi(x) p(x) dx \simeq \frac{1}{T} \sum_{\tau}^{T}\phi(\mathbf{x}^\tau), \textrm{where } \mathbf{x}^\tau \sim p(x)
\end{equation}
where $\mathbf{x}^\tau$ dictates a vector of points at position $\tau$ on a mesh.

Although difficult to effectively generate random samples from $\mathbf{x}\sim p(x)$, we can easily describe a probabilistic refinement of $\mathbf{x}$, described by the conditional distribution $q(\mathbf{x}^{l+1}|\mathbf{x}^{l})$. We consider repeated sampling from $q(\mathbf{x}^{l+1}|\mathbf{x}^{l})$, a markov chain, as the conditional probability distribution of future states of the process depends only upon the present state, not on the sequence of events that preceded it (markov property). This combined approach is usually referred to as  a Markov chain Monte carlo (MCMC) method. 

There are multiple conditional distributions, $q(x^{l+1}|x^{l})$ which can model $p(x)$. We utilise the gibbs process to determine the specific order we update our sampled value:
\begin{equation}
x^{l+1}_i \sim p(x^l_1, ... x^l_{i-1}, x^l_{i+1}, ..., x^l_n)
\end{equation}
Thus for each term $i$ of $x$ in turn, we sample a new value from the conditional distribution of $x_i$ given all other variables. It can be shown, that this will eventually generate dependent samples from the joint distribution p(x). Gibbs sampling reduces the task of sampling from a joint distribution, to sampling from a sequence of univariate conditional distributions.

\subsection{TrueSkill Ranking}
Originally introduced as a modification of the Elo chess ranking system, \cite{elo1978rating}. TrueSkill Ranking \cite{dangauthier2008trueskill} allows a comparative analysis of players skill given series of matches in which those individuals compete with a win or loss. Implemented on a large scalable factor graph, the model describes the probability of player 1 beating player 2 in some game, based on their  respective skills $w_{i}$. These skills are modelled by a multivariate distribution $P(\mathbf{w})$ an can be updated using Gibbs Sampling. 

Application of Bayesian inference to tasks such as TrueSkill Ranking have demonstrated a mean drift.  This originates from an observation that all data are comparative, thus all deduced latent variables are comparative by nature. After multiple iterations of Gibbs sampling, asymmetric models may continue to increase values of $x_{I}$ whilst sustaining the same comparative values, this results in a mean drift whilst the variance stays constant. A solution to mean drift is the ability to fix a given value $x_{I}$, thus anchoring all other values relative. 

\section{Model} 

In this section, we introduce how we model a stop-and-search scenario before exploring the interpretations.

\subsection{Model Proposal} We assume an individual $i$, has a latent variable $C_{i}$ which expresses their \textit{inherent criminality}, independent of if they are subject to a stop-and-search. We suppose the individual also has an additional attribute, their ethnicity $k_{i}$. An individual is subject to a stop-and-search if and only if $C_{i}+\beta_{k_{i}} +\mathcal{N}(0,\alpha)> 0$, where $\beta_k$ is their \textit{racial bias} dependent on ethnicity $k$, $\mathcal{N}$ is a normal distribution and $\alpha$ is a noise-term. Upon a search, an individual's \textit{true criminality} is uncovered, with an individual being found in possession if $C_{i} + \mathcal{N}(0,\gamma )>0$, where $\gamma$ is a further predetermined noise-term. Given a set of stop-and-search outcomes $\mathbf{X}$ with related ethnicity annotations, deduce $p(\beta_{k}|X)$.

In the situation where $\beta_k$ is equal for all values of $k$, we could make some claim that no racial bias' occurs within policing. On the other-hand, comparatively large values of $\beta$ would suggest a strong ethnically dependent bias disproportionately affecting some groups. 

\subsection{Choice of Priors}

Our choice of priors express the assumptions of the model and thus have severe ramifications on our outcomes. We produce two variations of our model with the following priors:

\textbf{Independent Bias} - We assume that $C$ and  $\beta_{k}$ are independent events. We model the \textit{inherent criminality} as $C_{i}\sim\mathcal{N}(0,1)$ and \textit{racial bias} as $B_{k_{i}}\sim\mathcal{N}(\mu_k, \sigma_k).$

We interpret this prior to suggest that Criminality is universally distributed amongst a diverse population, with the modal individual on the threshold of being discovered in the possession of something illegal. As noted within Section 2.4, a fixed mean anchors the relative bias. 

\textbf{Dependent Bias} - We assume that $C$ and  $\beta_{k}$ are dependent events. Thus we model these as a joint distribution $(C_{i}, \bm{\beta_{i}})\sim\mathcal{N}(\bm{\mu},\bm{\sigma})$. We model the \textit{inherent criminality} as $C_{i}\sim \int P(C,\bm{\beta}) d \bm{\beta} = \mathcal{N}(0,1)$ and \textit{racial bias} as $B_{k_{i}}\sim \int P(C,\bm{\beta}) dC =\mathcal{N}(\mu_k, \sigma_k)$. 

One can interpret this prior to suggest that each persons inherent criminality and ethnic bias has some correlation. Thus if an individual has a higher inherent criminality they may also subject to higher racial bias.

\textbf{Free Criminality} - We assume that $C$ and  $\beta_{k}$ are dependent events. Thus we model these as a joint distribution $(C_{i}, \bm{\beta_{i}})\sim\mathcal{N}(\bm{\mu},\bm{\sigma})$.  We place no further restrictions on the model.

\begin{figure*}[b!]
\centering
\includegraphics[width=0.85\textwidth]{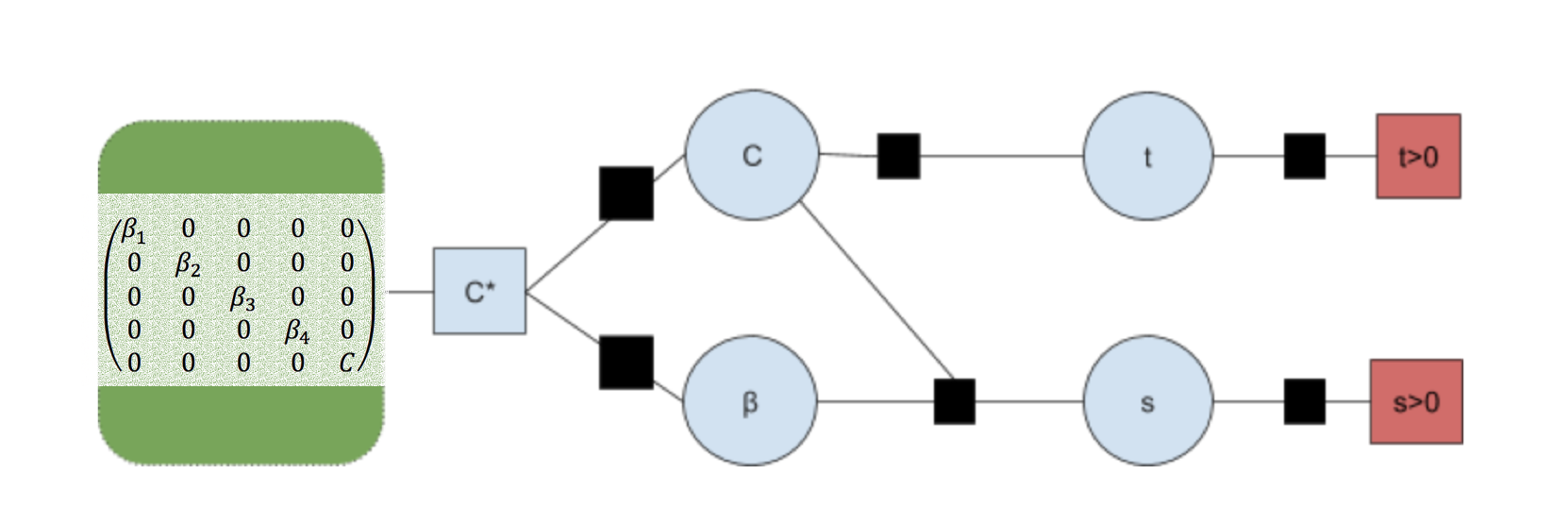}
\caption{A representation of the Factor Graph}

\end{figure*}

\subsection{Analytic Solution}We firstly attempt to derive a distribution for the racial bias conditioned upon our data $p(\beta_{k}|X)$, and taking the prior as the Independent Bias, $p(\theta|\alpha) = p(\beta_{k}) p(C)$. We take the noise $\alpha = \gamma =1$ and outcome as $X = \{0,1\}$ defined in Section 3.1 for ease of calculation:

\begin{equation}
p(S|C,\beta_{k}) = \mathcal{N}(S;C+\beta_{k},1)
\end{equation}
\begin{equation}
p(T|S,C) \approxprop  H(S)\mathcal{N}(T;C,1)
\end{equation}
\begin{equation}
p(X|T) =\delta(X-H(T))
\end{equation}
where, $H(x)$ refers to the Heaviside step function ($H(x) = 0, \forall x<0$).

We are motivated to introduce the Heaviside function in Eq.9 to moderate inspection of $T$. True criminality only be evaluated if an individual is actually stopped ($S>0$), thus is $S<0$, we should not be able to sample $T$. Thus $H(S)$ acts as a switch (for if an individual is not stopped but inspects their T value). A consequence of this introduction is a normalisation constant $Z=\int P(T|S,C)dSdTdC$. 

Applying Bayes Rule, for a single stop-and-search $i$, we evaluate the posterior as:
\begin{equation}
p(\beta_k|\mathbf{X},C)=\frac{p(\mathbf{X}|C, \beta)p(\beta_{k}) p(C)}{\int \int p(\mathbf{X}|C, \beta)p(\beta_{k}) p(C)d\beta_{k}dc}
\end{equation}

Expanding the likelihood function, $p(\mathbf{X}|C, \beta)$ and combining with Eq 6-8:
\begin{equation}
\int \int p(\mathbf{X}|T)p(T | S,C)p(S | C,\beta_{k})dSdT
\end{equation}

\begin{equation}
p(\mathbf{X}|C,\beta_{K})= \begin{cases}
 \Phi(C)\Phi(C+\beta_{k}) &\mbox{if } X = 1  \\
 \Phi(-C)\Phi(C+\beta_{k}) &\mbox{if } X \equiv 0
 \end{cases}.
\end{equation}
where $\Phi$ is the cumulative distribution function defined of a normal distribution $\mathcal{N}(0,1)$.

This has severe implications for our solution, as the likelihood function forces $C$ and $\beta$ to be correlated. Thus the posterior does not factorise. Furthermore, the posterior is no longer a Gaussian density function (due to the mixed cumulative distribution function). Thus we conclude that the current model does not have a closed analytic form. 

\subsection{Graphical Model:}
Given the inability to form an analytic solution, we turn to approximating the problem as a graph. We start by re-introducing two latent ($S_{i}, T_{i}$) and two observable($\Sigma_{i}, \tau_{i}$) variables:
\begin{equation}
S_{i} \sim \mathcal{N}(S | C_{i}+\beta_{k_{i}},\alpha)
\end{equation}
\begin{equation}
\Sigma_{i} = \textrm{sign}(S_{i})
\end{equation}
\begin{equation}
T_{i} \sim \mathcal{N}(T | C_{i},\gamma) 
\end{equation}
\begin{equation}
\tau_{i} =  \textrm{sign}(T_{i})
\end{equation}
where $S_{i}$ and $T_{i}$ are the \textit{stopping susceptibility} and \textit{true crimainlity} (resp.) of an individual, $i$. 

This new structure lets us form a factor graph (Figure 1). We note that unlike the formulation of the Analytic solutions we are able to remove the dependency of $T$ on $S$. This is motivated for two reasons: 1) Graphical models which include loops are not guaranteed to converge, and so removing one loop increases the ability to find a stable equilibria. 2) As a path still exists between $T$ and $S$, we are guaranteed that updating values of $S$ will still effect outcomes of $T$. 

The model requires no assumption of prior, as we can store both Independent Bias and Dependent Bias models in the same structure, $C*$ (a multivariate Gaussian distribution). How these priors affect our model is how we update of values of the variance during inference. We present our implementation of Gibbs sampling below.
\begin{algorithm}[ht]
   \caption{Gibbs Sampling for the Ethnic Bias Model}
   \label{alg:example} 
\begin{algorithmic}
   \STATE {\bfseries Input:} Network \{network structure g , latent variables S, T, \}, prior distribution $p(C^{0})$, tuple of observations/training examples $\{(\Sigma,\tau, k)\}_{n}$
   \STATE {\bfseries Output:} posterior distribution $p(\beta_k|\mathbf{X},C)=p(C*)$
   \STATE{\bfseries Initialise:} Sample $(c,b_{k})\sim p(C^0) = p(C,\beta_{k})$
   \FOR{ $iterations$}
   \FOR{ $\{(\Sigma,\tau, k)\}_{n}$ in training examples}
   \STATE update $S \sim \frac{1}{Z_1} \delta(1-\Sigma) \mathcal{N}(S | c+b_{k},\alpha)$
   \STATE update $T \sim \frac{1}{Z_2} \delta(1-\tau) \mathcal{N}(T | c,\gamma)$
   
   \STATE update $ \sigma(C) +=  1$, $ \sigma(\beta_{k}) +=  1$
   \IF {Dependent Bias}
   \STATE update $ \sigma(C,\beta_k) += 1$,  $ \sigma(\beta_K,C) +=  1$

  \ENDIF
   \STATE update $ \mu(C) += -\tau (T - c)$
   \STATE update $ \mu(\beta_K) +=  S-\beta_{k} - C$

   \ENDFOR
  \STATE update $\sigma(C^*) = \sigma(C^0)+\sigma(C) $
   \STATE update $\mu(C^*)= \sigma(C^*)$$(\sigma(C^0)^{-1}\mu(C^0) +\sigma(C)^{-1}\mu(C))$
   \ENDFOR
\end{algorithmic}
\end{algorithm}

We note that although simpler to explain the gibbs method, we optimise certain steps using message passing, we include an implementation in the appendix. 

Finally, as mentioned within Section 2.5, models were latent variables are only defined relative to each other, and observables are defined using inequalities can lead to mean drift. To compensate for this in large models, we reduce the mean vector $\mu=(\beta_{1},...,\beta_{k},C)$ by subtracting $C$ and scale our variance $\sigma$ matrix by $\sigma(C)$, thus keeping the relative values but reducing the chances of model explosion.
\section{Experiments}
In this section we apply the approaches defined in Section 2 and Section 3. We firstly provide a statistical analysis of the data to establish some ground truths, before applying a naive approach to further reinforce these. We then infer the values for $\beta_{k}$ under various model constraints. We implement a mixed method of gibbs sampling and message passing, with a compute time of 141 iterations/second on 2.8GHz CPU. Finally we demonstrate the extensibility of the model, inspect results on a Territorial level and also highlight the versatility of our modelling different situations.

\subsection{Data}
We aggregate the data from the Home Office Police, stop-and-search database. The dataset has over 16,224 stop and searches reported, each stop comes with a ethnicity labelled by the Office instigating and also self-identified label. We inspect 4 main datasets, \textbf{National}, \textbf{Augmented}, \textbf{London Met} and \textbf{Charges}. 

 \textbf{National}: The national dataset, is a combination of all 47 police forces within the UK. All responses are grouped into two categories \{Not Guilty, Guilty\}. Below we include an breakdown in terms of ethnicity and guilty/not guilty.
\begin{center}
\includegraphics[width=0.5\textwidth]{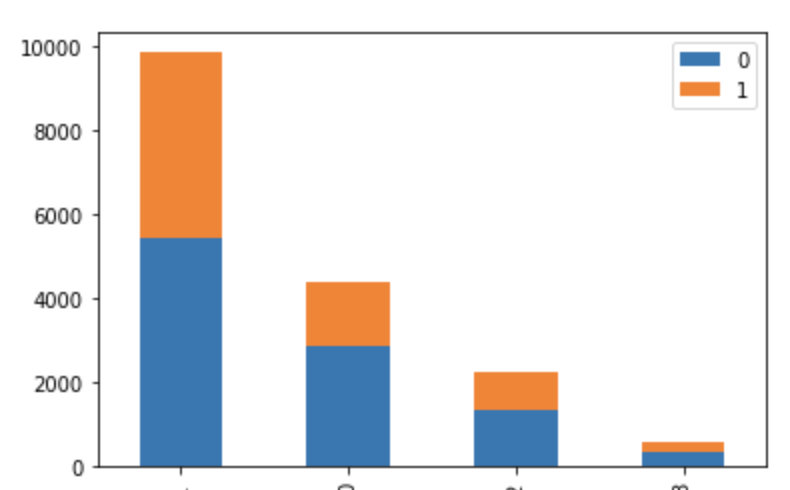}
\begin{tabular}{| l |c|c|r |}
\hline
  Ethnicity & Stopped Nationally & Guilty (\%)&Column \\  \hline
  White & 9374 & 44.89&0 \\ \hline
  Black & 4168 & 35.09&1 \\ \hline
  Asian& 2146 & 39.80&2 \\ \hline
  Other/Mixed & 536 & 40.88&3 \\ \hline
\end{tabular}
\end{center}
 \textbf{Augmented}: We also create an augmented dataset where we control the variance in numbers stopped. This will provide useful in distinguishing how bias our model is to the number of stops it is able to observe.
\begin{center}
\begin{tabular}{| l |c|r |}
\hline
  Ethnicity & Stopped Nationally & Guilty (\%) \\ \hline
  White & 1000& 45.60\\ \hline
  Black & 1000 & 33.60  \\ \hline
  Asian& 1000 & 40.20 \\ \hline
  Other/Mixed & 1000 & 41.60\\ \hline
  \end{tabular}
\end{center}
\textbf{Charges:} To inspect the severity of charges we create the Charges dataset. Response to being found in possession is provided in over 14 classes of responses. These classes are not mutually exclusive given the different naming conventions of different Police Forces. For our previous data sets, we simplify our results to \{Lenient, Severe\}.

We define lenient as \{'Khat or Cannabis Warning", "Local resolution", "Community resolution", "A no further action disposal", "Suspected substances seized - No further action"\}. Whilst severe is defined as all other responses to being found guilty. Thus our dataset only includes responses to those found guilty. Below we provide a breakdown of percentage receiving less severe charges to total guilty charges by ethnicity. 

\begin{center}
\includegraphics[width=0.5\textwidth]{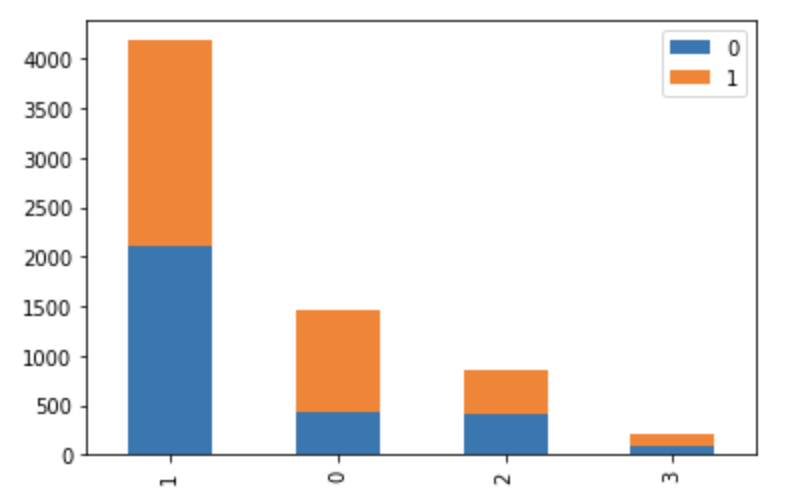}
\end{center}
\begin{center}
\begin{tabular}{| c |c|c |r|}
\hline
  Ethnicity & Total Guilty & Less Severe(\%) &Column\\ \hline
  White & 4183& 50.29&0\\  \hline
  Black & 1466& 31.50 &1 \\ \hline
  Asian& 860 & 47.21&2 \\ \hline
  Other/Mixed & 220 & 38.46 &3\\ \hline
\end{tabular}
\end{center}

\textbf{London Met:} We also choose to look at a territorial level and in particular examine the London Metropolitan Police. This data provides a more balanced distribution of ethnicities and race. 
\begin{center}
\includegraphics[width=0.5\textwidth]{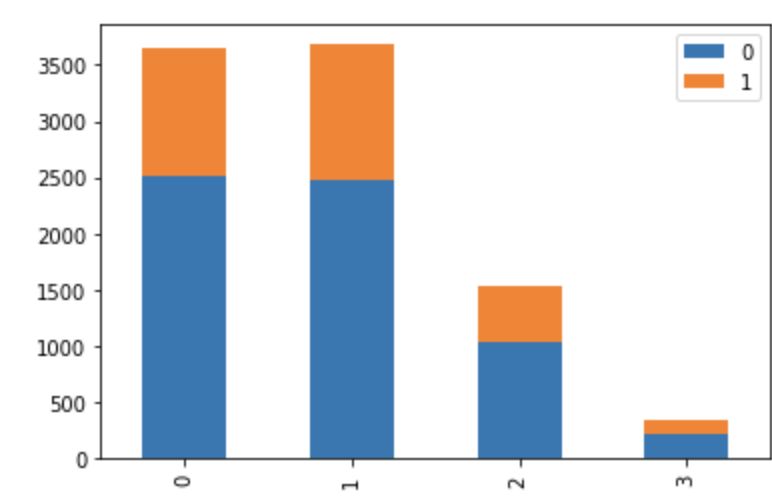}
\begin{tabular}{| l |c|c |}
\hline
  Ethnicity & Stopped by London Met & Guilty (\%) \\ \hline
  White & 3679& 32.62\\ \hline
  Black & 3657 & 31.42  \\ \hline
  Asian& 1536 & 30.33 \\ \hline
  Other/Mixed & 351 & 36.46 \\ \hline
\end{tabular}
\end{center}
\subsection{Naive Approach}
We first attempt to understand the data better and gain some first order approximations to how we should expect our $\beta_{K}$ to be ranked.  We formulate our problem as such:

There are a set of Criminalities $(C_{1}, C_{\beta_{1}} ... C_{\beta_{k}})$. During a stop and search, and individual of ethnicity $k_{I}$ if an individual is found guilty, $C_{1}>C_{\beta_{1}} $ and if found not guilty, then  $C_{1}<C_{\beta_{1}} $. 

We use gibbs sampling over 500 iterations to produce the following rank,
\begin{center}
\includegraphics[width=0.5\textwidth]{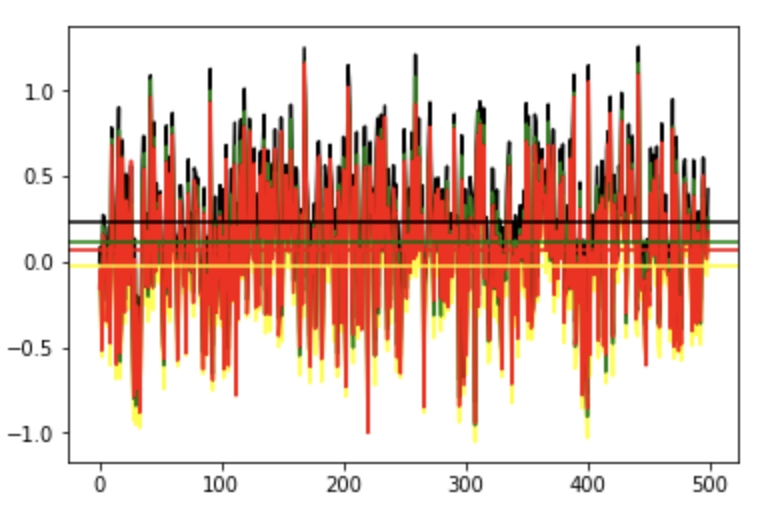}
\end{center}

Note within our game a higher $C_{\beta_{k}}$ relates to a higher chance of being stopped but being found not guilty. Although different $C_{\beta_{k}}$ do not directly play against each other, they do play against the common player $C_{1}$, from this we find the following ranking
\begin{center}
\begin{tabular}{| c |c|c |}
\hline
  Rank & Ethnicity & Mean/Skill   \\ \hline
  1 & Black & 0.21879219 \\ \hline
  2 & Asian &  0.10181663 \\ \hline
  3& Other/Mixed & 0.05889284 \\ \hline
  4& White & -0.03291718 \\ \hline
  5 & Criminality & -0.17103525 \\ \hline
  
\end{tabular}
\end{center}
Comparing these results with the \% Guilty, this first order approximation looks reasonable, however the closeness of classes Asian and Other/Mixed is seemingly exaggerated. Finally from the graph above, we can see that the variance dominates the Other/Mixed class, this is due to lower number of stops compared to other biases.

\subsection{Independent Bias - Gibbs Sampling}
We next the model proposed within Section 3.4, using Gibbs sampling, the model settles into the follow values. We report the mean and variance for each individual bias after over 500 iterations of the model and we believe the values have reasonably converged. 
\begin{center}
\begin{tabular}{| c |c|c |c|}
\hline
 Rank & Ethnicity & Bias Mean &  Bias Variance \\ \hline
  1 &Other/Mixed & 2.57986335&1.73549436\\ \hline
  2 & Asian &  1.28414313&7.51716549 \\ \hline
  3&  Black & 1.10097326&3.88335607\\ \hline
  4& White & 0.87199493& 1.73549436 \\ \hline
\end{tabular}
\end{center}

From these observations we note that our model is not matching the ground truths. To further explore this, inspect values during the iterations of the Gibbs sampling (displayed below). It is clear that the amount of data has a direct effect on the variance of each bias term.
\begin{center}
\includegraphics[width=0.5\textwidth]{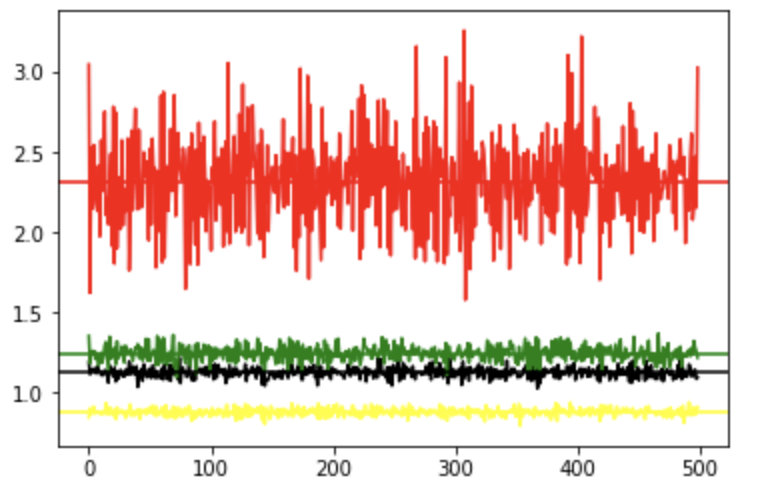}
\end{center}
With our rankings resembling the inverse of size of available data, this suggests that the model is asymmetric and having more instances of stop and searches may affect the bias calculation (an method for updating a particular value may be heavily affected by the size of available data).

We clarify this by using an augmented dataset (where the number of stops is controlled). We find that the model is now matches the percentage guilty, with each distribution having the same variance ($\sigma = 3.3.99401$), however the mean values are closer together than previously. 
 \begin{center}
\includegraphics[width=0.5\textwidth]{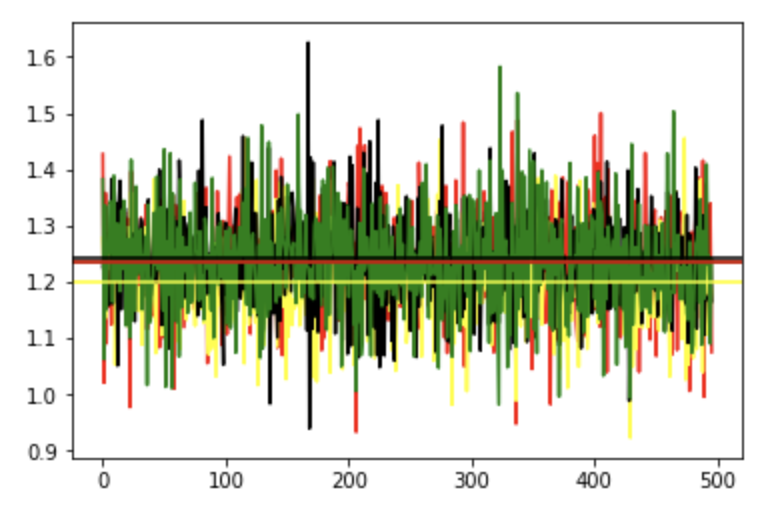}
\end{center}

\subsection{Dependent Bias - Gibbs Sampling}
We now apply the model again, this time using the Dependent Prior. Immediately we note that the covariance term has restricted the variance of each of the bias distributions. Furthermore we note the rankings now agree with the ground truth.
\begin{center}
\includegraphics[width=0.5\textwidth]{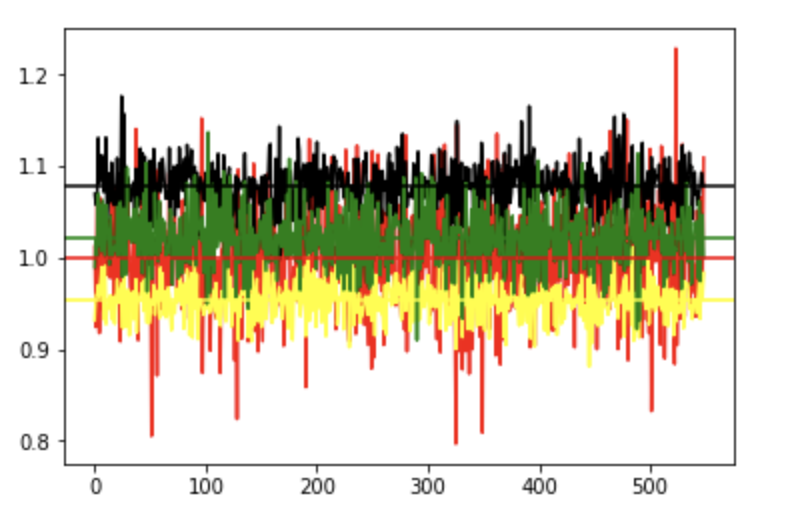}
\end{center}
\begin{center}
\begin{tabular}{| c |c|c |c|}
\hline
  Rank & Ethnicity & Mean &Variance\\ \hline
  1 &Black & 1.07857592&1.00068161\\ \hline
  2 & Asian &  1.02163574&0.99945829\\ \hline
  3&  Other/Mixed & 0.99905217& 0.99780932\\ \hline
  4& White &0.95345612& 1.0003046\\ \hline
\end{tabular}
\end{center}

As we can see the variances are significantly more entwined than in the Independent Bias model. To provide the counterfactual, we also tested our model on the augmented dataset. We find mean values are with a $\pm0.1$ of their un-augmented counterparts and a fixed variance of $\sigma =3.3.99401$, rankings continued to match the ground truths.

\subsection{Free Criminality}
We include only the gibbs plot for the model. What we found most interesting about this plot, is that this shows that the model is numerically unstable. This is expected as the prior does not constrain the values in any way. We note that in this method our criminality stabilises in a distribution $C= -0.223$. Over 1 Million iterations we found no stable ranking.
\begin{center}
\includegraphics[width=0.5\textwidth]{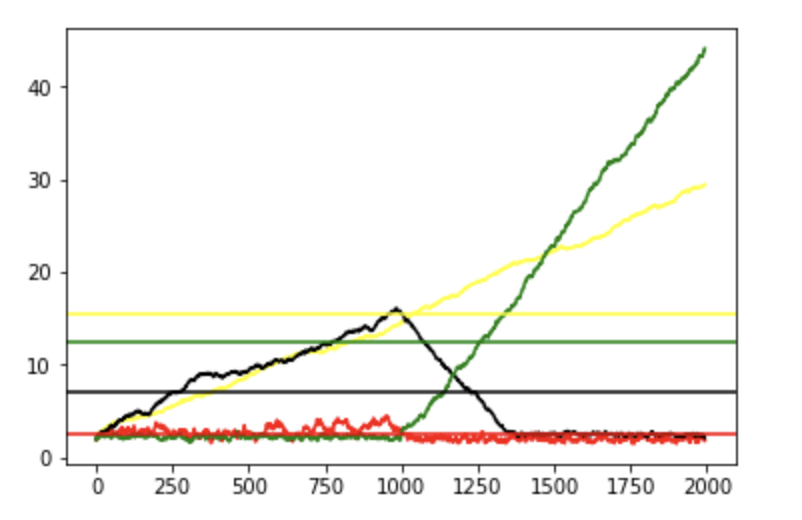}
\end{center}

\subsection{Possible Extensions}
We demonstrate the versatility of this model with two restrictions. We apply the dependent prior model to the Charges dataset, demonstrating the ability for the model to function on less data. In this description fo the model, our threshold is transfer from being stopped to being found guilty, whilst now our value of "truth" event, $T$ is now defined to be if the charge is severe (as opposed to lenient). We include the biases from the Severity dataset: 
\begin{center}
\begin{tabular}{| c |c|c |c|}
\hline
  Rank & Ethnicity & Mean & Variance\\ \hline
  1 &Asian &  0.79589927&1.003478\\ \hline
  2 & Black &  0.74368696&1.00204095\\ \hline
  3&  White &0.73769407&1.0007155\\ \hline
  4& Other/Mixed & 0.72061036&1.01356524 \\ \hline
\end{tabular}
\end{center}

\section{Discussion}
In this section, we discuss the limitations and restrictions of this model. 

\textbf{Main Objective:}
The main objective of this paper was to evaluate \textit{if} racial biases exist, which we believe we have proved to be true. This was easily validated by a comparison of mean values of each distribution.

\textbf{Number of Datapoints:}
From section 4.4 we realise that our the independent prior model is distorted by the number of stop-and-searches provided. Number of searches as a quantity should have no weight on a model, but should be taken in the context of the size of the general population. For example, Black searches make up approximately 25\% of total searches is a meaningless statement, however given the context that Black population is 13\%, this shows a disproportionate amount of searches. Unfortunately as only a small number of the total population is subject to stop and searches, our model has no concept of the entire population.

This example highlights a larger problem with the dataset, and the asymmetric information in which we only are aware of people who have been stopped, not those who the Police do not deem worth stopping. 

\textbf{Validation:}
It is difficult to provide clear validation of the models. In the experiment section we establish ground truths against \% Guilty, however this hardly provides much additional insight. Using the generative model and distributions, we reserve 10\% of the data for a test set, which we tested we generate a score of 36\%, 42\% and 56\% (Independent, Dependent and Free Prior respectively). We find it alarming that a model which is allowed to explode (and does), still performs better than those we restrict. This suggests that the model itself needs to be limited in different ways.

A final validation is to compare with current literature on the subject. Analysis by Bowling and Philips \cite{bowling2007disproportionate}, identify key statistics for the probability of being stopped given race. Interestingly the data they analyses also includes Traffic stops, which are not included in this data set. Whilst again, we find common ranking of bias as ours, the generative claims of our model do not match.

Our weak ability to predict suggest there are significant other factors to incorporate (such as Age, Gender and Location). We believe that these are easily factored in via the Eq.12, such that $S_{i} \sim \mathcal{N}(S | C_{i}+\Sigma_j^J F_j,\alpha)$, where $F_j$ is an additional factor we wish to monitor.

\section{Conclusion}
In this paper we have discussed a series of Bayesian tools for application to the UK Stop-and-Search database. We propose a model and identify a combination of gibbs sampling and message passing as an optimised form to produce latent priors.

We find conclusively that racial bias' are present within stop-and-searches and provide a tool for analysing this at Force by Force basis. We however find it difficulty to quantify measures of how biased some forces are compared to others. 

We believe that with the consideration of extra factors and the availability of more data we can further quantify and understand how to more effectively calculate the biases introduced in this paper.

\bibliography{main.bib}
\bibliographystyle{icml2017}

\end{document}